\documentclass[12pt]{article}
\usepackage{amsmath,amssymb,epsf,epsfig,amsthm,bm,graphicx}
\usepackage[dvips]{color}

\begin{document}

\title{ \flushright{\small CECS-PHY-08/17; AEI-2008-089} \\
\center{On the equivalence of the SO(2,1) gauged WZW form and
Liouville theory}
}\author{\textbf{Andr\'{e}s Anabal\'{o}n}$^{1,2,3}$\thanks{%
andres-dot-anabalon-at-uai-dot-cl} \\
\textbf{\textit{$^{1}$}}{\small Departamento de Ciencias, Facultad
de Artes Liberales,}\\
{\small   Facultad de Ingenier\'{\i}a y Ciencias,
Universidad Adolfo Ib\'{a}\~{n}ez, Vi\~{n}a del Mar, Chile.}\\
 \textbf{\textit{$^{2}$}}{\small Max-Planck-Institut
f\"{u}r
Gravitationsphysik, Albert-Einstein-Institut, }\\
{\small Am M\"{u}hlenberg 1, D-14476 Golm, Germany.}\\
\textbf{\textit{$^{2}$}}{\small Centro de Estudios Cient\'{\i}ficos (CECS)
Casilla 1469 Valdivia, Chile.}\\
}
\maketitle

\begin{abstract}
It is proved that Liouville theory and the two dimensional SO(2,1) gauged
Wess-Zumino-Witten term are on-shell equivalent. This shed light on a
possible higher dimensional generalization of the former theory.
\end{abstract}

The study of three dimensional gravity has shed light on the existence of a
fiber bundle structure in most of the known gravitational theories. In
particular, three dimensional Einstein gravity turns out to be defined as a
theory of flat connections \cite{Achucarro:1987vz}, and thus derivable from
a three dimensional Chern-Simons (\textbf{CS}) action. Indeed, defining the
connection as being valued in, for instance, the three dimensional AdS group
$SO(2,2)$:

\begin{equation}
\mathcal{A}=\frac{1}{2}\omega ^{AB}J_{AB}+\frac{e^{A}}{l}P_{A},\qquad
\mathcal{F}=\frac{1}{2}\left[ R^{AB}+\frac{1}{l^{2}}e^{A}e^{B}\right] J_{AB}+%
\frac{1}{l}T^{A}P_{A},
\end{equation}%
where $J_{AB}$ are Lorentz rotations, $P_{A}$ are anti de Sitter
translations, $R^{AB}=d\omega ^{AB}+\omega _{C}^{A}\omega ^{CB}$, $%
T^{A}=de^{A}+\omega _{B}^{A}e^{B}\equiv De^{A}$, and $A=0,1,2$. Three
dimensional Einstein AdS gravity is nothing but the statement that
\begin{equation}
\mathcal{F}=0.  \label{eq}
\end{equation}%
Although this was discovered in the context of the $OSp(\left. p\right\vert
2;%
\mathbb{R}
)\otimes $ $OSp(\left. q\right\vert 2;%
\mathbb{R}
)$ extension of $SO(2,2)$ three dimensional gravity \cite{Achucarro:1987vz},
it has become useful in the construction of extended supergravities (for an
extensive catalog of the maximally supersymmetric gauged supergravities in
three dimensions see \cite{Nicolai:2000sc}) and also turns out to be
relevant in the holographic context. When a CS theory is defined in a
manifold with a boundary, it defines a chiral WZW theory \cite{Moore:1989yh,
Elitzur:1989nr} on its boundary. However in the case of $SO(2,2)$, gravity
it is necessary to properly define the meaning of asymtotically AdS. This
amount to impose Brown-Henneaux asymptotic conditions \cite{Coussaert:1995zp}%
, or even weaker ones \cite{Rooman:2000zi}, reducing the WZW theory to the
Liouville theory\footnote{%
It has also been proved that is not necessary to resort to the CS
formulation to induce Liouville theory as the asymptotic dynamics of three
dimensional AdS gravity \cite{Carlip:2005tz}.}.

Its connection with CS gravity suggests that Liouville theory itself should
admit a description in terms of a purely topological object, which should
inherit the structure from the CS form. This is actually the
case,classically (and no quantum consideration will occupy us in what
follows), as will be shown in this paper.

The even dimensional sibling of CS forms, is the gauged WZW form \cite%
{Witten:1983tw, AlvarezGaume:1983cs}, which in two dimensions defines the
following action:%
\begin{equation}
S(h,\mathcal{A})=\kappa \int_{\Sigma }\frac{1}{3}\left\langle \left(
h^{-1}dh\right) ^{3}\right\rangle -\kappa \int_{\partial \Sigma
}\left\langle \left( \mathcal{A}-h^{-1}dh\right) h^{-1}\mathcal{A}%
h\right\rangle -\kappa \int_{\partial \Sigma }\left\langle \mathcal{A}%
h^{-1}dh\right\rangle \text{,}  \label{1}
\end{equation}%
where the brackets stand for an invariant tensor, $h$ for a group element, $%
\mathcal{A}$ for a connection and the wedge products are omitted to simplify
the notation. It is invariant under the adjoint action of the group:

\begin{equation}
\mathcal{A}\rightarrow \mathcal{A}^{g}=g^{-1}\mathcal{A}g+g^{-1}dg\text{,}%
\qquad h\rightarrow g^{-1}hg\text{.}  \label{gauget}
\end{equation}%
The action (\ref{1}) is usually accompanied with a kinetic term for $h$.
However, in this case the kinetic term is omitted since it would not allow
to extrapolate these results to higher dimensions, where the non-linear
sigma model kinetic term is not conformal invariant. The previously quoted
facts mark an important difference with the interesting and useful Liouville
embedding in WZW theories known previously \cite{Forgacs:1989ac}.

In this case $SO(2,1)$ is the relevant gauge group and the generators satisfy

\begin{equation}
\left[ P_{a},P_{b}\right] =\mp m^{2}J_{ab}\text{,}\qquad \left[ J_{ab},P_{c}%
\right] =-P_{b}\eta _{ac}+P_{a}\eta _{bc}\text{,}
\end{equation}%
where $a=0,1$, $\eta _{ab}(-,+)$. Here it is interesting to note that a
redefinition of the generators allows to consider negative or positive $%
m^{2} $ terms and therefore both cases are covered in this approach.

The connection and the curvature are

\begin{equation}
\mathcal{A}=\frac{1}{2}\omega ^{ab}J_{ab}+me^{a}P_{a}\text{,}
\end{equation}

\begin{equation}
\mathcal{F}=\frac{1}{2}\left( R^{ab}\mp m^{2}e^{a}e^{b}\right) J_{ab}+\left(
de^{a}+\omega _{c}^{a}e^{c}\right) mP_{a}.
\end{equation}%
The fact that $\omega ^{ab}$ is a Lorentz connection implies that the action
of the covariant derivative, $D=d+\omega $, on the invariant tensor $\eta
_{ab}$ identically vanishes\footnote{%
This can be relaxed considering $\omega ^{ab}$ not valued in the Lorentz
algebra \cite{Cacciatori:2005wz}.}, $D\eta \equiv 0.$This, in turn, implies
that the metric defined as

\begin{equation}
g_{\mu \nu }=e_{\mu }^{a}e_{\nu }^{b}\eta _{ab},  \label{met}
\end{equation}%
satisfies

\begin{eqnarray}
\partial _{\lambda }g_{\mu \nu } &=&D_{\lambda }g_{\mu \nu }=\left(
D_{\lambda }e_{\mu }^{a}\right) e_{\nu }^{b}\eta _{ab}+e_{\mu }^{a}\left(
D_{\lambda }e_{\nu }^{b}\right) \eta _{ab}\equiv e_{\alpha }^{a}\Gamma
_{\lambda \mu }^{\alpha }e_{\nu }^{b}\eta _{ab}+e_{\mu }^{a}\Gamma _{\lambda
\nu }^{\beta }e_{\beta }^{b}\eta _{ab}  \notag \\
&=&\Gamma _{\lambda \mu }^{\alpha }g_{\nu \alpha }+\Gamma _{\lambda \nu
}^{\beta }g_{\mu \beta }.  \label{MC}
\end{eqnarray}%
Where the use of the relation
\begin{equation*}
D_{\lambda }e_{\mu }^{a}\equiv e_{\alpha }^{a}\Gamma _{\lambda \mu }^{\alpha
},
\end{equation*}%
can be simply regarded as a redefinition of the LHS object. Eq. (\ref{MC})
is nothing but the well known definition of metric compatibility. Thus, when
the metric (or equivalently the vielbein) is invertible, the condition $%
de^{a}+\omega _{c}^{a}e^{c}=0$ identifies $\Gamma $ as the Levi-Civitta
connection. Here it should be remarked that once the vielbein is taken to be
part of a gauge connection one is confronted with the fact that gauge
transformations can map configurations where the vielbein is invertible to a
non-invertible one, something that has been matter of controversy \cite%
{Witten:1988hc, Witten:2007kt}. However, once an invertible vielbein has
been chosen, small gauge transformations do not affect its invertibility.
Indeed, small AdS translations, $\delta e^{a}=D\lambda ^{a}$, can be
identified as diffeomorphisms

\begin{equation*}
\delta g_{\mu \nu }=\delta \left( e_{\mu }^{a}e_{\nu }^{b}\eta _{ab}\right)
=\nabla _{\mu }\lambda _{\nu }+\nabla _{\nu }\lambda _{\mu },
\end{equation*}%
where $\nabla _{\mu }\lambda _{\nu }=\partial _{\mu }\lambda _{\nu }+\Gamma
_{\mu \nu }^{\alpha }\lambda _{\alpha }$. At the end of the paper the
invertibility of the vielbein in gauged WZW forms gravities and Chern-Simons
gravities is contrasted.

Now, let us show that the field equations that follow from (\ref{1}) implies
that the conformal factor of the spacetime metric (\ref{met}) satisfies the
Liouville equation. To this end we set the invariant tensor as the Killing
metric: $\left\langle J_{01}J_{01}\right\rangle =1$, $\left\langle
J_{01}P_{a}\right\rangle =0$, $\left\langle P_{a}P_{b}\right\rangle =\eta
_{ab}$.

The field equations for $\mathcal{A}$ reads

\begin{equation}
\left\langle G_{A}\left( h^{-1}\mathcal{A}h+h^{-1}dh-h\mathcal{A}%
h^{-1}-hdh^{-1}\right) \right\rangle =0\text{,}  \label{2}
\end{equation}%
and the field equations for $h$ are

\begin{equation}
\left\langle G_{A}\left( h^{-1}\mathcal{F}h+\mathcal{F-}\frac{1}{2}\left[
\left( \mathcal{A}-h^{-1}\mathcal{A}h-h^{-1}dh\right) ,\left( \mathcal{A}%
-h^{-1}\mathcal{A}h-h^{-1}dh\right) \right] \right) \right\rangle =0\text{,}
\label{3}
\end{equation}%
where $G_{A}$ stands for all the generators.

Let us consider first (\ref{2}): to solve them, it is necessary to select a
coordinate chart in the group manifold. Since any group element of $SO(1,2)$
can be locally written as the adjoint action of the group on the exponential
of a Cartan subalgebra, there is the freedom to select $h$ as

\begin{equation}
h=p\exp (\frac{1}{2}\lambda ^{ab}J_{ab})p^{-1}\text{,}  \label{4}
\end{equation}%
where $p$ is some group element that makes (\ref{4}) a good coordinate
chart. In this way, it is possible to recognize that most of the field
content of $h$ can be eliminated by selecting the gauge transformation in (%
\ref{gauget}) as $g=p$.

Thus, we pick

\begin{equation}
h=\exp (\frac{1}{2}\lambda ^{ab}J_{ab})\text{.}  \label{gf}
\end{equation}%
In this gauge the field equations (\ref{2}) are:

\begin{eqnarray}
\delta e^{1} &:&e^{0}\sinh (\lambda ^{01})=0\text{,}  \label{a1} \\
\delta e^{0} &:&e^{1}\sinh (\lambda ^{01})=0\text{,} \\
\delta \omega ^{ab} &:&d\lambda ^{ab}=0\text{,}  \label{a5}
\end{eqnarray}

We see that, (\ref{a5}) implies that $\lambda ^{ab}$ is constant. Moreover,
equation (\ref{a1}) and the invertibility of the vielbein implies:

\begin{equation}
\sinh (\lambda ^{01})=0\text{,}
\end{equation}%
When the vielbein is invertible the most general solution to the field
equations is:

\begin{equation}
\lambda ^{01}=0\Longrightarrow h=1\text{.}  \label{h}
\end{equation}

This, in turn, solve all the field equations (\ref{2}) and simplifies the
remaining ones (\ref{3}) to the form:

\begin{equation}
\mathcal{F}=0\text{,}
\end{equation}%
As discussed before the equation of motion $de^{a}+\omega _{c}^{a}e^{c}=0$
set the connection to be the Levi-Civitta connection and $\left( R^{ab}\mp
m^{2}e^{a}e^{b}\right) =0$ is just:

\begin{equation}
R=\pm 2m^{2},  \label{L1}
\end{equation}%
or in the conformal gauge, $g_{\mu \nu }=e^{2\phi }\tilde{g}_{\mu \nu }$:

\begin{equation}
\Box \phi \pm 2m^{2}e^{2\phi }-\frac{1}{2}R\left( \tilde{g}_{\mu \nu
}\right) =0,  \label{LE}
\end{equation}

This complete the demonstration of the classical equivalence of (\ref{1})
and Liouville theory.

\textbf{On the vielbein invertibility.}

It was very early realized that a subsector of a Chern-Simons gauge
theory
it was classically equivalent to three dimensional general relativity \cite%
{Achucarro:1987vz}. Therefore it was  argued that the corresponding
quantum theory should include all the possible configurations; in
particular the sector of vanishing connection, $e=0=\omega$
 \cite{Witten:1988hc}. Later it was remarked in \cite{Witten:2007kt}
that the introduction of such singularities is not needed in the sum
over complex structures of two dimensional quantum gravity. Although
some singularities are allowed to compactify the moduli space of the
sum over two dimensional geometries they are of much milder nature
than $e=0=\omega ,$ in the sense that there is still a geometrical
interpretation of them (the structure of the moduli space of two
dimensional quantum gravity is nicely discussed in
\cite{D'Hoker:1988ta}).

However, already at the classical level there is a problem with the
Chern-Simons gauge theory as a theory of gravity. Namely that the
configuration $\mathcal{A}=0$ has no meaning in terms of Riemannian
geometry. Thus if the Chern-Simons form is taken as the one that
defines the Lagrangian of three dimensional gravity one has to a
priori exclude the configurations where the vielbein is not
invertible. However the gauge symmetry of the theory imply the
equivalence of globally well defined configurations such as anti de
Sitter spacetime to the configuration $\mathcal{A}=0$ (on the
contrary, the BTZ black hole presents a global obstruction to
achieve this non-geometrical configuration). Therefore the
functional space of connections has no gauge invariant separation
between the sectors of invertible and non-invertible vielbeins. In
the gauged WZW gravity the situation is different. Once the gauge
has been fixed as in (\ref{gf}) an invertible vielbein can not be
mapped to a non-invertible one by the residual gauge freedom. Thus,
once the gauge is fixed there is a natural gauge invariant
separation of the functional space between the sectors of invertible
vielbeins and non-invertible ones. The gauge freedom of the theory
is therefore compatible with the geometrical interpretation of it.

Finally, it is worth to remark that the rather peculiar form (\ref{1}) of
writing this conformal field theory allows a higher dimensional
generalization. Indeed, in every even dimensional spacetime it is possible
to write down a gauged WZW form. This has been investigated in four
dimensions \cite{Anabalon:2007dr}, however we have imposed different
constraints in the field content of these theories in order to reproduce
General Relativity. At the light of the results presented here, one should
consider the four dimensional generalization with no restriction, with the
hope of find a four dimensional analog of the Liouville theory.

\textbf{Acknowledgments.}

The author would like to thank Joaquim Gomis, Mokhtar Hassaine, Marc Magro,
Illarion Melnikov, Julio Oliva, Ricardo Troncoso, Steve Willison and Jorge
Zanelli many discussions and guide in the literature that was essential to
understand this problem. This work is supported by the grant No. 3080024
from FONDECYT (Chile) and for the Alexander von Humboldt foundation
(Germany). The Centro de Estudios Cient\'{\i}ficos (CECS) is funded by the
Chilean Government through the Millennium Science Initiative and the Centers
of Excellence Base Financing Program of Conicyt. CECS is also supported by a
group of private companies which at present includes Antofagasta Minerals,
Arauco, Empresas CMPC, Indura, Naviera Ultragas and Telef\'{o}nica del Sur.

\end{document}